\documentclass[12pt,a4j]{article}
\usepackage[dvipdfmx]{color, graphicx}
\usepackage{amsmath,enumerate, amsfonts, ulem, url}
\usepackage{natbib}
\usepackage{authblk}
\usepackage{multirow}
\usepackage{here}
\usepackage[subrefformat=parens]{subcaption}
\begin{document}
\baselineskip 18pt
	\begin{center}
		{\LARGE\textbf{Classification from Positive and Biased Negative Data with Skewed Labeled Posterior Probability }}
	\end{center}
	\begin{center}
		{\large Shotaro Watanabe$^1$ and Hidetoshi Matsui$^2$\footnote{E-mail: \texttt{hmatsui@biwako.shiga-u.ac.jp}}}
	\end{center}
	
	\begin{center}
		\begin{minipage}{14cm}
			{
				\begin{center}
					{\it {\footnotesize 
							$^1$Graduate School of Data Science, Shiga University \\
							$^2$Faculty of Data Science, Shiga University \\
							1-1-1, Banba, Hikone, Shiga, 522-8522, Japan.
					}}
					
				\end{center}
				\vspace{1mm} 
				{\small {\bf Abstract:}
					The binary classification problem has a situation where only biased data are observed in one of the classes. In this paper, we propose a new method to approach the positive and biased negative (PbN) classification problem, which is a weakly supervised learning method to learn a binary classifier from positive data and negative data with biased observations. We incorporate a method to correct the negative impact due to skewed confidence, which represents the posterior probability that the observed data are positive. This reduces the distortion of the posterior probability that the data are labeled, which is necessary for the empirical risk minimization of the PbN classification problem. We verified the effectiveness of the proposed method by numerical experiments and real data analysis.
				}
				
				\vspace{3mm}
                }
		\end{minipage}
	\end{center}
	
\section{Introduction}
The purpose of binary classification is to identify whether a sample belongs to a positive or a negative class, and this classification is applicable to various fields, such as life science, materials science, and marketing. In binary classification problems, conventional supervised classification learns a classifier using fully positive (P) data and fully negative (N) data. We call this problem PN classification. In contrast, weakly supervised machine learning has been widely studied in recent years, including positive and Unlabeled (PU) classification, which learns a classifier using P data and unlabeled (U) data (Elkan $\&$ Noto 2008; du Plessis et al. 2014; du Plessis et al. 2015; Kiryo et al. 2017), semi-supervised classification (Chapelle et al. 2006; Sakai et al. 2017), and noisy-label learning (Natarajan et al. 2013; Shi et al. 2018; Zhang et al. 2019). \par
In this paper, we consider the situation where the majority of the N data are not available. In this situation, it is difficult to collect complete N data, but it may be easier to collect only a small set of biased N data. We call these data biased negative (bN) data. As applications for analyzing P and bN data, Li et al. (2010) and Fei $\&$ Liu (2015) both tackled the problem in the context of text classification. Li et al. (2010) mentioned that bN data may have a negative impact on classification, and completely truncated bN data and P and U data were used for classification. Fei $\&$ Liu (2015) considered the situation in which collecting unbiased U data would be difficult, and trained a classifier using only P and bN data. However, this method is specialized for text classification because it relies on the domain knowledge; it uses an effective similarity measure to evaluate the similarity between documents. In contrast to these two studies, Hsieh et al. (2019) proposed PUbN classification, the method for learning from P, U, and bN data without requiring specific domain knowledge for fitting arbitrary classifiers (ranging from linear to deep models). However, their method assumes the use of U data and it is difficult to learn from only P and bN data. When U data are not available, it is difficult to estimate the distribution of the entire feature set, which makes it difficult to learn the classifier. In this paper, we extend PUbN classification and propose PbN classification, which trains a classifier using only P and bN data even when U data and domain knowledge are not available. \par
In PbN classification, one problem is the posterior probability for each instance observation cannot be obtained accurately. As an approach to this problem, we use the idea of positive confidence (Pconf) classification. Pconf classification is a learning method in which a binary classifier is learned from only P data with a confidence that represents the positive posterior probability, without N or U data (Ishida et al., 2018). Pconf classification gives a "skew" due to the fact that the confidence cannot be obtained explicitly. Here we follow Shinoda et al. (2020) and refer to the deformation of confidence and other parameters by bias as ``skew". We incorporate a measure that corrects the negative effect of the skew of the confidence of Pconf classification in PbN classification. We call this method adjusted PbN classification. We verified the effectiveness of the proposed method by applying it to the analysis of synthetic data and benchmark data. \par
This paper is organized as follows. In Section 2, we review the problem setup for PN and Pconf classification. In Section 3, we first present the problem setup of PbN classification, and then show how to adjust the posterior probability that each instance is the skewed observation. In Section 4, we compare the performance of adjusted PbN classification with that of PN and naive PbN classification using synthetic data and benchmark data, and describe how we verified the effectiveness of the method. Finally, we present our conclusions in Section 5.
\section{Problem Formulation and Existing Method}
In this section, we review the problem setup for PN and Pconf classification.
\subsection{PN Classification}
Let $x\in\mathbb{R}^d, y\in\{+1, -1\}$ be a $d$-dimensional feature vector and a label, respectively, and suppose they follow an unknown probability density function $p(x,y)$. Let $p_P(x)=p(x|y=+1)$ and $p_N(x)=p(x|y=-1)$ be distributions of features with positive and negative labels, respectively, and let $\pi=p(y=+1)$ be a class prior. Let $g : \mathbb{R}^d\rightarrow\mathbb{R}$ be a binary classifier and $\ell : \mathbb{R}\rightarrow\mathbb{R}_+$ be a loss function. We denote $R^+_P(g)=E_{x\sim p_P(x)}[\ell(g(x))]$ and $R^-_N(g)=E_{x\sim p_N(X)}[\ell(-g(x))]$ as an expected false negative rate and an expected false positive rate, respectively. Then the PN classification risk is expressed as
\begin{align}
R(g)=\pi R^+_P(g)+(1-\pi)R^-_N(g).
\end{align}
We determine $g$ by minimizing the classification risk (1). In practice, the minimizer of the classification risk (1) cannot be calculated directly because it includes the expectation for the loss function. However, if the P data set $\mathcal{X}_P=\{x^P_i\}^{n_P}_{i=1}$ and the N data set $\mathcal{X}_N=\{x^N_i\}^{n_N}_{i=1}$ are observed independently from $p_P(x)$ and $p_N(x)$, respectively, the classification risk (1) can be approximated as 
\begin{align}
\hat{R}(g)=\pi\hat{R}^+_P(g)+(1-\pi)\hat{R}^-_N(g), \nonumber
\end{align}
where $\hat{R}^+_P(g)=\tfrac{1}{n_P}\sum^{n_P}_{i=1}\ell(g(x^P_i))$ and $\hat{R}^-_N(g)=\tfrac{1}{n_N}\sum^{n_N}_{i=1}\ell(-g(x^N_i))$ are empirical losses of false negative and false positive rates, respectively.
\subsection{Pconf Classification}
Pconf classification is an approach for binary classification from only P data under the condition that neither N nor U data are observed. In this situation, it is difficult to evaluate $R_N^-$ in the classification risk. Ishida et al. (2018) expressed the Pconf classification risk as  
\begin{align}
R(g)=\pi\left[R^+_P(g)+R^-_P\left(\frac{1-r}{r}g\right)\right],
\end{align}
where $R^-_P(g)=E_{x\sim p_P(X)}[\ell(-g(x))]$ and $r=p(y=+1|x)$ is the confidence that the observed data are P. In Pconf classification, the classifier can be trained from only $\mathcal{X}_P=\{x^P_i\}^{n_P}_{i=1}$, and then the classification risk can be approximated as 
\begin{align}
\hat{R}(g)=\pi\left[\hat{R}^+_P(g)+\hat{R}^-_P\left(\frac{1-r}{r}g\right)\right], \nonumber
\end{align}
where, $\hat{R}^-_P(g)=\tfrac{1}{n_P}\sum^{n_P}_{i=1}\ell(-g(x^P_i))$. \par
To obtain the confidence $r$, it is necessary to evaluate marginal density $p(x)$ or $p(x|y=-1)$ directly, or to use logistic regression from fully P and N data as a classifier. However, it is difficult to consider the situation where they are known in real-world applications. Therefore, we need to estimate the confidence $r$ with the help of the domain knowledge by experts, but this may cause bias and skew of the confidence. Shinoda et al. (2020) assumed a false negative rate of the classification as a prior knowledge, and modified the confidence by minimizing the squared difference between the false negative rate and the empirical false negative rate. 
\section{Proposed Method}
In this section, we first propose a classification risk for PbN classification, and then propose a method for correcting the skew of the posterior probability that each instance is observed, that occurs in PbN classification. 
\subsection{PbN Classification}
Unlike PN learning, PbN learning considers the situation where only some biased negative data are observed rather than all negative data are observed in a binary classification problem. To solve this problem, we introduce a latent variable $s$ that returns $+1$ for the observed P and bN data and $-1$ for the unobserved N data. The class prior probability $\pi$ is assumed to be known in this paper. Let $\rho=p(y=-1, s=+1)$, $p_{bN}(x)=p(x|y=-1, s=+1)$, and $p_{s=+1}(x)=p(x|s=+1)$, further, let $\sigma=p(s=+1|x)$ be the probability that the data are observed. Let $R^-_{bN}(g)=E_{x\sim p_{bN}(x)}[\ell(-g(x))]$ and $R^-_{s=+1}(g)=E_{x\sim p_{s=+1}(x)}[\ell(-g(x))]$, then if $\sigma\neq0$, the PbN classification risk is expressed as
\begin{align}
R(g)=\pi R^+_P(g)+\rho R^-_{bN}(g)+(\pi+\rho)R^-_{s=+1}\left(\frac{1-\sigma}{\sigma}g\right).
\end{align}
The derivation of (3) is given in Appendix A.\par
Let $\mathcal{X}_{bN}=\{x^{bN}_i\}^{n_{bN}}_{i=1}, \mathcal{X}_{s=+1}=\{x^{s=+1}_i\}^{n_{p}+n_{bN}}_{i=1}$ be observed independently from $p_{bN}(x)$ and $p_{s=+1}(x)$, respectively. Then, the classification risk (3) can be approximated as 
\begin{align}
\hat{R}(g)=\pi \hat{R}^+_P(g)+\rho\hat{R}^-_{bN}(g)+(\pi+\rho)\hat{R}^-_{s=+1}\left(\frac{1-\sigma}{\sigma}g\right), \nonumber
\end{align}
where $\hat{R}^-_{bN}(g)=\tfrac{1}{n_{bN}}\sum^{n_{bN}}_{i=1}\ell(-g(x^{bN}_i))$ and $\hat{R}^-_{s=+1}=\tfrac{1}{n_p+n_{bN}}\sum^{n_p+n_{ bN}}_{i=1}\ell(-g(x^{s=+1}_i))$. In this case, as in the case of Pconf classification, $\sigma$ cannot be obtained explicitly because it depends on $p(x)$. However, if we use bN data, we can obtain a pseudo $p(x)$ as $p_{\mathrm{bias}}(x)=\pi p(x|y=+1)+(1-\pi)p(x|y=-1, s=+1)$, and unlike the Pconf classification, we can obtain the skewed $\sigma$ directly as follows: 
\begin{align}
\tilde{\sigma}=\frac{p(s=+1)p(x|s=+1)}{p_{\mathrm{bias}}(x)}.
\end{align}
\subsection{Adjusted PbN Classification}
To correct the skew of $\sigma$, we incorporate a measure to correct the skew of the confidence level in Pconf classification derived by Shinoda et al. (2020). We consider the following classification risk with hyperparameter $k$ ($0<k<\infty$).
\begin{align}
\hat{R}_{\mathrm{bias}}(g)=\pi \hat{R}^+_P(g)+\rho\hat{R}^-_{bN}(g)+(\pi+\rho)\hat{R}^-_{s=+1}\left(\frac{1-\tilde{\sigma}^k}{\tilde{\sigma}^k}g\right). \nonumber
\end{align}
The hyperparameter $k$ can be selected by cross-validation if there is a validation set containing P and N data. However, it is difficult to use the cross-validation because there are only P and bN data. To solve this problem, we assume that the following false negative rate for classification is known as prior knowledge.
\begin{align}
\phi = \int_{\{x: g(x)<0\}}p(x|y=+1)dx. \nonumber
\end{align}
Under this assumption, we can select the optimal hyperparameter $k^*$, which minimizes the difference between the known false negative rate and the empirical false negative rate, by minimizing the following squared error
\begin{align}
k^* = \mathrm{arg}\underset{k}{\mathrm{min}}\left(\frac{1}{n_p}\sum^{n_p}_{i=1}\ell_{01}(g(x_i))-\phi\right)^2, \nonumber
\end{align}
where $\ell_{01}(z)=(1-\mathrm{sign}(z))/2$ is the 0-1 loss. The selected $k^*$ is used to express the adjusted risk for the PbN classification as
\begin{align}
\hat{R}_{\mathrm{bias}}(g)=\pi \hat{R}^+_P(g)+\rho\hat{R}^-_{bN}(g)+(\pi+\rho)\hat{R}^-_{s=+1}\left(\frac{1-\tilde{\sigma}^{k^*}}{\tilde{\sigma}^{k^*}}g\right). \nonumber
\end{align}
In practice, the assumption that the false negative rate $\phi$ is known is not realistic. However, as shown in the numerical experiments in Section 4.1.3, the classification accuracy of the proposed method is not sensitive to the choice of $\phi$.
\section{Experiments}
This section describe how we validated the performance of our proposed method with experiments using synthetic data and benchmark data.
\subsection{Synthetic Data Experiments} 
\subsubsection{Set up}
In these experiment, we generated a two-dimensional training dataset, a validation dataset, a test dataset, and a dataset for false negative rate estimation from P and N data generated independently. The training dataset contained 500 P samples and 100 bN samples, and the validation dataset contained 500 P samples. Furthermore, the test dataset and the dataset for false negative rate estimation contained 500 P samples and 500 N samples. We compared the classification accuracy of the adjusted PbN classification with that of the naive PbN classification and the ordinary PN classification. For the naive PbN classification and PN classification, the training dataset and the validation dataset were combined as the training dataset. \par
For all methods, we used the linear model $g(x)=a^Tx+\beta$ for the binary classifier and the logistic loss $\ell_L(z)=\mathrm{log}(1+e^{-z})$ for the loss function. Furthermore, the stochastic gradient descent was applied for optimization to construct the learner. The candidate of the hyperparameter $k$ was selected from $\{0.3, 0.5, 0.7, 1.0, 1.5, 2.0, 4.0\}$. Note that $\hat{\sigma}$ less than 0.01 was rounded up to 0.01 for optimization stability. \par
The P data were generated from a Gaussian distribution with a mean vector $[0,0]^\top$ and identity covariance matrix. In contrast, the N data were assumed to be generated from a mixture of four Gaussian distributions whose means were different from each other and whose covariance matrices were all identity matrices. We considered two patterns according to the degree of overlap for the P and N data. For the pattern with the smaller overlap, the N data were distributed by the mixture of normal distributions with mean vectors $[1.0, 1.0]^\top$, $[1.5, 1.5]^\top$, $[2.0, 2.0]^\top$, and $[2.5, 2.5]^\top$; conversely, for the pattern with a larger overlap, the N data were distributed by a mixture of distributions with mean vectors $[2.0, 2.0]^\top$, $[3.0, 3.0]^\top$, $[4.0, 4.0]^\top$, and $[5.0, 5.0]^\top$. By applying the proposed method to the analysis of these datasets, we examined the difference in accuracy depending on the degree of overlap of the classes. We also considered two cases depending on which data were actually observed in a biased manner from the N data; specifically, the N data were observed from only one of the four components, and the N data were observed partially from each of four components with probability $[0.25, 0.25, 0.25, 0.25]$/$[0.40, 0.10, 0.35, 0.15]$/$[0.15, 0.40, 0.10, 0.35]$/$[0.35, 0.15, 0.40, 0.10]$. \par
To summarize the above, we conducted numerical experiments under the following four situations. 
\begin{quote}
 \begin{itemize}
  \item \textbf{Situation 1. }The degree of the class overlap is large and one of the four components of the N data is actually observed
  \item \textbf{Situation 2. }The degree of the class overlap is small and one of the four components of the N data is actually observed
  \item \textbf{Situation 3. }The degree of the class overlap is large and the observed N data are partially obtained from each of the four components of the N data
  \item \textbf{Situation 4. }The degree of the class overlap is small and the observed N data are partially obtained from each of the four components of the N data
 \end{itemize}
\end{quote}
In each of these settings, we calculated the classification accuracy by naive PbN classification, adjusted PbN classification, and PN classification.
\subsubsection{Experiments with four different situations}
We report the mean and standard deviation of the classification accuracy for the 10 trials of Situation 1 and Situation 2 in Table 1 and Table 2, respectively, where $\mu_{\mathrm{bn}}$ is the mean of the bN data, and $\hat{\phi}$ is the mean and standard deviation of the estimated false negative rate over 10 trials. A.PbN and N.PbN represent the adjusted PbN classification and the naive PbN classification, respectively. The adjusted PbN classification is significantly better than the PN classification when the class overlap is large, and significantly better than the naive PbN classification when the class overlap is small. \par
\begin{table}[t]
\caption{Mean and standard deviation of the classification accuracy for 10 trials with Situation 1. The best method based on the 5$\%$ t-test and equivalent methods are shown in bold. \\}
\centering
\begin{tabular}{l|c|c|c|c} \hline
$\mu_{\mathrm{bn}}$ & A.PbN & N.PbN & PN & $\hat{\phi}$ \\ \hline
$[1.0, 1.0]^T$ & \pmb{\bf $85.31\pm0.82$} & $83.39\pm1.15$ & $83.04\pm0.97$ & \multirow{4}{*}{$10.50\pm1.20$} \\
$[1.5, 1.5]^T$ & \pmb{$86.09\pm0.54$} & \pmb{\bf $85.91\pm0.79$} & $83.35\pm1.15$ &  \\
$[2.0, 2.0]^T$ & \pmb{$85.70\pm0.65$} & \pmb{$86.01\pm1.27$} & $82.19\pm0.79$ &  \\
$[2.5, 2.5]^T$ & \pmb{$85.73\pm1.02$} & \pmb{$85.78\pm0.79$} & $79.26\pm1.10$ &  \\ \hline
\end{tabular}
\end{table}
\begin{table}[t]
\caption{Mean and standard deviation of the classification accuracy for 10 trials with Situation 2. The best method based on the 5$\%$ t-test and equivalent methods are shown in bold. \\}
\centering
\begin{tabular}{l|c|c|c|c} \hline
$\mu_{\mathrm{bn}}$ & A.PbN & N.PbN & PN & $\hat{\phi}$ \\ \hline
$[2.0, 2.0]^T$ & \pmb{$95.81\pm0.57$} & $93.72\pm0.38$ & \pmb{$96.14\pm0.40$} & \multirow{4}{*}{$2.55\pm0.71$} \\
$[3.0, 3.0]^T$ & \pmb{$96.10\pm0.65$} & $92.18\pm1.10$ & $94.40\pm0.77$ &  \\
$[4.0, 4.0]^T$ & \pmb{$95.88\pm0.74$} & $84.20\pm2.19$ & $91.26\pm1.07$ &  \\
$[5.0, 5.0]^T$ & \pmb{$95.37\pm1.02$} & $76.86\pm1.74$ & $87.10\pm0.75$ &  \\ \hline
\end{tabular}
\end{table}
Next, the mean and standard deviation of the classification accuracy in 10 trials with Situation 3 and Situation 4 are shown in Table 3 and Table 4, respectively. The adjusted PbN classification is significantly better than the PN classification when the class overlap is large, and significantly better than the naive PbN classification when the class overlap is small. These results show that the classification accuracy of the proposed method is stable regardless of the class overlap and the shape of the bN data. \par
The decision boundaries in Situations 1--4 are shown in Figure 1 (a)--(d), respectively. The blue, green, and orange points represent the P data, observed bN data, and unobserved N data, respectively. The decision boundaries for naive PbN classification, adjusted PbN classification, and PN classification are shown in yellow, red, and purple, respectively. From these figures, we can see that the naive PbN classification is strongly influenced by the P data and the PN classification is strongly influenced by the bN data compared to the proposed method. 
\begin{table}[t]
\caption{Mean and standard deviation of the classification accuracy for 10 trials with Situation 3. The best method based on the 5$\%$ t-test and equivalent methods are shown in bold. \\}
\centering
\begin{tabular}{l|c|c|c|c} \hline
bN & A.PbN & N.PbN & PN & $\hat{\phi}$ \\ \hline
$[0.25, 0.25, 0.25, 0.25]$ & \pmb{\bf $86.18\pm1.05$} & \pmb{$85.92\pm0.85$} & $82.50\pm1.01$ & \multirow{4}{*}{$10.58\pm1.34$} \\
$[0.40, 0.10, 0.35, 0.15]$ & \pmb{$86.08\pm0.99$} & \pmb{$86.08\pm1.11$} & $82.79\pm1.35$ &  \\
$[0.15, 0.40, 0.10, 0.35]$ & \pmb{$86.64\pm1.05$} & $86.24\pm1.30$ & $82.49\pm1.21$ &  \\
$[0.35, 0.15, 0.40, 0.10]$ & \pmb{$86.04\pm1.03$} & \pmb{$85.94\pm1.11$} & $83.26\pm1.50$ &  \\ \hline
\end{tabular}
\end{table}
\begin{table}[htbp]
\caption{Mean and standard deviation of the classification accuracy for 10 trials with Situation 4. The best method based on the 5$\%$ t-test and equivalent methods are shown in bold. \\}
\centering
\begin{tabular}{l|c|c|c|c} \hline
bN & A.PbN & N.PbN & PN & $\hat{\phi}$ \\ \hline
$[0.25, 0.25, 0.25, 0.25]$ & \pmb{$96.21\pm0.79$} & $86.93\pm1.45$ & $95.19\pm0.44$ & \multirow{4}{*}{$2.52\pm0.78$} \\
$[0.40, 0.10, 0.35, 0.15]$ & \pmb{$96.38\pm0.38$} & $88.80\pm1.12$ & $95.74\pm0.55$ &  \\
$[0.15, 0.40, 0.10, 0.35]$ & \pmb{$95.46\pm0.55$} & $86.89\pm1.96$ & $94.13\pm0.72$ &  \\
$[0.35, 0.15, 0.40, 0.10]$ & \pmb{$95.82\pm0.65$} & $89.66\pm1.83$ & $95.06\pm0.52$ &  \\ \hline
\end{tabular}
\end{table}
\begin{figure}[htbp]
 \begin{minipage}[b]{1.0\linewidth}
  \centering
  \includegraphics[keepaspectratio, scale=0.32]{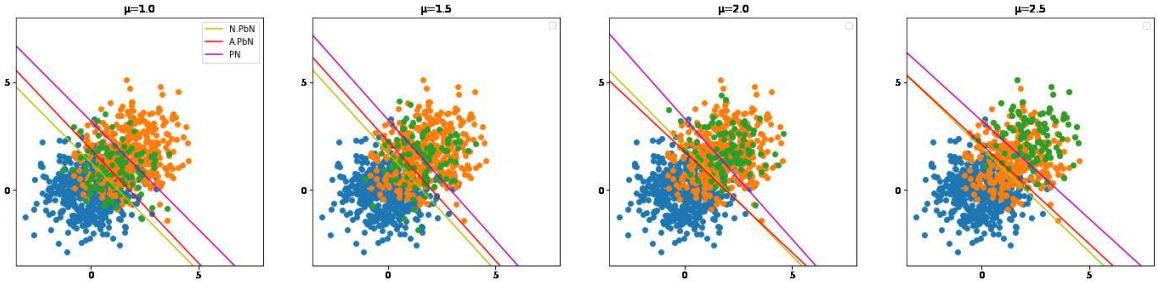}
  \subcaption{Decision boundary of Situation 1}\label{poly04}
 \end{minipage} \\
 \begin{minipage}[b]{1.0\linewidth}
  \centering
  \includegraphics[keepaspectratio, scale=0.32]{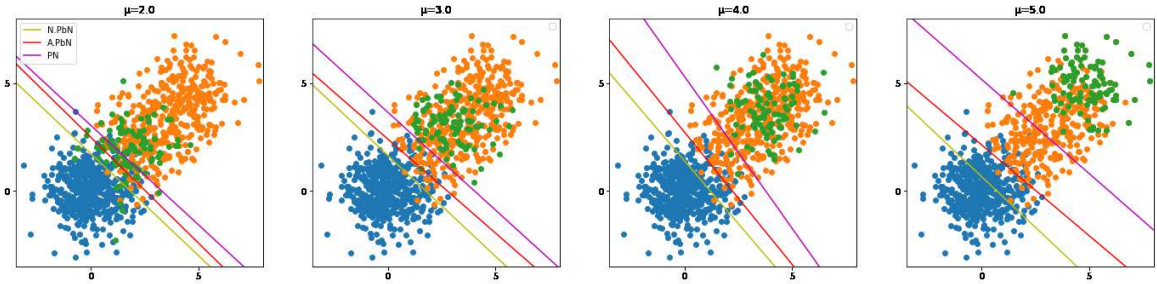}
  \subcaption{Decision boundary of Situation 2}\label{poly06}
 \end{minipage} \\
 \begin{minipage}[b]{1.0\linewidth}
  \centering
  \includegraphics[keepaspectratio, scale=0.32]{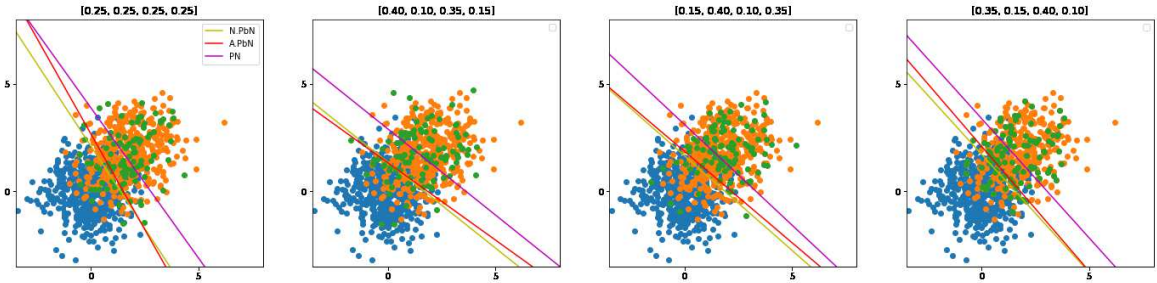}
  \subcaption{Decision boundary of Situation 3}\label{poly08}
 \end{minipage}\\
 \begin{minipage}[b]{1.0\linewidth}
  \centering
  \includegraphics[keepaspectratio, scale=0.32]{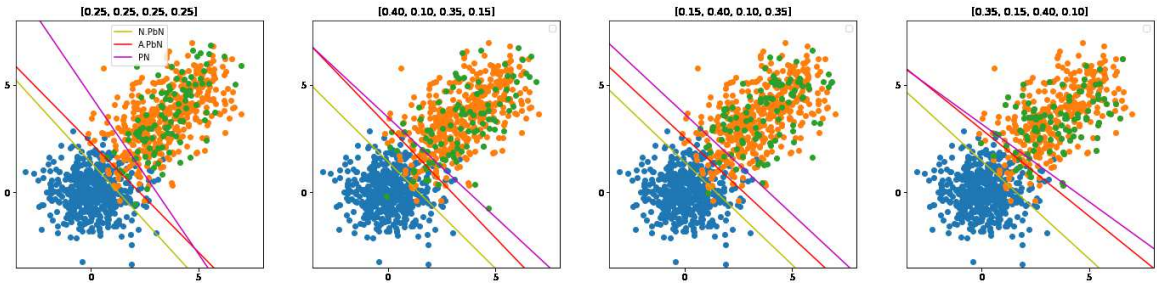}
  \subcaption{Decision boundary of Situation 4}\label{poly12}
 \end{minipage}
 \caption{Decision boundaries of four situations.}\label{reg_poly}
\end{figure}
\subsubsection{Impact of Estimation Error of $\phi$}
In the experiments described in Section 4.1.2, we assumed that the false negative rate $\phi$ can be estimated accurately, but in practice, this assumption does not necessarily hold. Therefore, we conducted some experiments with deviations for $\hat\phi$ to see how much the estimation error of $\phi$ affects the accuracy of the adjusted PbN classification. \par
We applied the adjusted PbN classification in the same setting as Situation 1 and Situation 2 described in Section 4.1.2, but with the correctly estimated values of $\hat{\phi}$ multiplied by $c\in\{0.5, 0.7, 1.3, 1.5\}$, which causes a deviation to the estimation. The mean and standard deviation of the classification accuracy in 10 trials are shown in Tables 5 and 6. When the degree of class overlap is large, in the case of $c=0.5, 0.7$, the effect of estimation error is large, but when the degree of class overlap is small, the effect of the estimation error is negligible unless $\hat{\phi}$ is underestimated. This result indicates that the proposed method is effective even in practical situations where $\phi$ cannot be estimated accurately, as long as the estimation error is evaluated to be large or the degree of class overlap is small.
\begin{table}[t]
\caption{Means and standard deviations of the classification accuracy in 10 trials of the adjusted PbN classification using $\hat{\phi}$ with the estimation error when the degree of class overlap is large. Based on a 5$\%$ t-test, results equivalent to those for $c=1.0$ are shown in bold. $\mu_{\mathrm{bn}}$ is the mean of the bN data. \\}
\centering
\begin{tabular}{l|c|c|c|c} \hline
$\mu_{\mathrm{bn}}$ & $c=0.5$ & $c=0.7$ & $c=1.3$ & $c=1.5$ \\ \hline
$[1.0, 1.0]$ & $79.23\pm2.65$ & $83.24\pm2.19$ & \pmb{$85.53\pm0.89$} & \pmb{$85.66\pm0.78$} \\
$[1.5, 1.5]$ & $81.44\pm1.67$ & $83.68\pm1.29$ & \pmb{$85.21\pm0.52$} & $85.14\pm0.47$ \\
$[2.0, 2.0]$ & $81.02\pm2.32$ & $84.04\pm1.57$ & \pmb{$85.80\pm1.74$} & \pmb{$85.00\pm1.82$} \\
$[2.5, 2.5]$ & $79.74\pm2.76$ & \pmb{$83.14\pm2.27$} & \pmb{$84.27\pm1.53$} & $82.74\pm1.69$ \\ \hline
\end{tabular}
\end{table}
\begin{table}[t]
\caption{Means and standard deviations of the classification accuracy in 10 trials of the adjusted PbN classification using $\hat{\phi}$ with the estimation error when the degree of class overlap is small. Based on a 5$\%$ t-test, results equivalent to those for $c=1.0$ are shown in bold. $\mu_{\mathrm{bn}}$ is the mean of the bN data. \\}
\centering
\begin{tabular}{l|c|c|c|c} \hline
$\mu_{\mathrm{bn}}$ & $c=0.5$ & $c=0.7$ & $c=1.3$ & $c=1.5$ \\ \hline
$[2.0, 2.0]$ & $89.67\pm1.79$ & $94.32\pm1.04$ & \pmb{$96.24\pm0.53$} & \pmb{$96.24\pm0.53$} \\
$[3.0, 3.0]$ & $90.61\pm1.38$ & \pmb{$95.14\pm1.48$} & \pmb{$95.76\pm0.45$} & \pmb{$95.76\pm0.45$} \\
$[4.0, 4.0]$ & $91.60\pm3.95$ & \pmb{$94.74\pm1.60$} & \pmb{$95.44\pm0.64$} & \pmb{$95.30\pm0.77$} \\
$[5.0, 5.0]$ & $89.91\pm2.71$ & \pmb{$95.67\pm0.70$} & \pmb{$95.44\pm0.48$} & \pmb{$95.44\pm0.48$} \\ \hline
\end{tabular}
\end{table}
\subsection{Benchmark Experiments}
We used the Wireless Indoor Localization Data Set, available from the UCI Machine Learning Repository (Lichman 2013). This dataset was collected for an experiment to locate indoor positions using wireless LAN signal strength. The dataset consists of four rooms numbered by 1--4, each with a sample size of 500 consisting of 7-dimensional feature vectors. Here, we created a dataset for binary classification with room 2 as a positive class and the other rooms as a negative class. The bN data were classified in two patterns: those obtained only from rooms 1, 3, and 4, and those obtained randomly from each room. This dataset was then divided into a training dataset, a validation dataset, a test dataset, and a dataset for false negative rate estimation. The training dataset contained 200 P samples and 100 bN samples, the validation dataset contained 100 P samples, and the test dataset and the dataset for false negative rate estimation respectively contained 100 P samples and 300 N samples.\par
We compared the classification accuracy of the adjusted PbN classification with that of the naive PbN classification and the ordinary PN classification. For all methods, the linear model $g(x)=a^Tx+\beta$ was used as the binary classifier, the logistic loss $\ell_L(z)=\mathrm{log}(1+e^{-z})$ was used as the loss function, and the stochastic gradient descent method was applied for optimization to construct the learner. For the estimation of $p(x|s=+1)$, $p(x|y=+1)$, and $p(x|y=-1, s=+1)$, which are necessary for the estimation of the probability $\sigma$ that the data are observed, we used a Gaussian kernel and kernel density estimation with a bandwidth of 0.1. The candidates of the hyperparameter $k$ were $\{0.5, 0.7, 0.9, 1.0, 1.2, 1.5, 2.0\}$, and $\hat{\sigma}$ less than 0.01 was rounded up to 0.01 for optimization stability. \par
The means and standard deviations of the correct answers in 100 trials are shown in Table 7, where $\hat{\phi}$ is the mean and standard deviation of the estimated false negative rate over 100 trials. The proposed method performed as well as or better than the existing methods in all cases.\par
\begin{table}[t]
\caption{Mean and standard deviation of the percentage of classification accuracy over 100 trials for the Wireless Indoor Localization Data Set when the bN data are obtained only from rooms 1, 3, and 4, and when the data are obtained randomly from each room. The best method based on the 5$\%$ t-test and equivalent methods are shown in bold. \\}
\centering
\begin{tabular}{l|c|c|c|c} \hline
$\mu_{\mathrm{bn}}$ & A.PbN & N.PbN & PN & $\hat{\phi}$ \\ \hline
Room 1. & \pmb{\bf $81.44\pm7.26$} & \pmb{$80.73\pm5.91$} & $76.56\pm5.37$ & \multirow{4}{*}{$3.66\pm6.53$} \\
Room 3. & \pmb{$98.77\pm0.72$} & $97.79\pm2.29$ & \pmb{$98.50\pm1.55$} &  \\
Room 4. & \pmb{$77.14\pm2.06$} & \pmb{$76.16\pm1.90$} & $70.83\pm1.88$ &  \\
random & \pmb{$97.02\pm1.73$} & \pmb{$96.80\pm1.76$} & $95.02\pm7.42$ &  \\ \hline
\end{tabular}
\end{table}
\section{Conclusion}
In this paper, we propose a new approach for the PbN classification problem under the condition that only positive data and some biased negative data are observed. In the proposed method, the risk cannot be estimated properly due to the skew of the probability that indicates whether each instance is actually observed. The skew is reduced by minimizing the squared error of the false negative rate and the empirical false negative rate. We confirmed that our method is superior to the existing classification methods through analyses of synthetic and benchmark datasets. We also found that under certain circumstances, even if there is a deviation in the estimation of the false negative rate, the effect on the classification accuracy is small. \par
In this study, we considered the case where U data are not available. In our work, we estimated the probability density of the data by the kernel density estimation. In this case, it was difficult to provide learning with on high-dimensional data, unlike PUbN classification, because PUbN classification treats P data and bN data as P data, and then solves the PU classification problem by applying a deep neural network with the sigmoid function for the output layer, to estimate the probability that the data are observed. The future work is to consider a method for the high-dimensional situation where it is difficult to estimate the probability density. Furthermore, we have to explore another method for hyperparameter selection that does not require the assumption that the false negative rate is known. 


\appendix
\def\thesection{\Alph{section}}
\section*{Appendix}
\section{Derivation of PbN Classification Risk} 
The second term of the classification risk (1) can be decomposed as follows:
\begin{align}
(1-\pi)R^-_N(g)&=(1-\pi)\int\ell(-g(x))p(x|y=-1)dx \nonumber \\
&=\int p(y=-1)p(x|y=-1)\ell(-g(x))dx \nonumber \\
&=\int p(x, y=-1)\ell(-g(x))dx \nonumber \\
&=\int \left[p(x, y=-1, s=+1)+p(x, s=-1)\right]\ell(-g(x))dx \nonumber \\
&=\int p(y=-1, s=+1)p(x|y=-1, s=+1)\ell(-g(x))dx \nonumber \\
&\quad + \int p(x, s=-1)\ell(-g(x))dx \nonumber \\
&=\rho R^-_{bN}(g) + \int p(x, s=-1)\ell(-g(x))dx.
\end{align}
because $p(x, s=-1, y=+1)=0$ and $p(x, s=-1)=p(x, s=-1, y=-1)$. Under the assumption that $\sigma = p(s=+1|x)\neq 0$, we obtain
\begin{align}
p(x,s=-1)+p(x,s=+1)&=p(x) \nonumber \\
&=\frac{p(x, s=+1)}{p(s=+1|x)}. \nonumber
\end{align}
Then,
\begin{align}
p(x, s=-1)&=\frac{p(x, s=+1)\left[1-p(s=+1|x)\right]}{p(s=+1|x)} \nonumber \\
&=\frac{1-\sigma}{\sigma}p(x, s=+1). \nonumber
\end{align}
Therefore, the second term in (4) can be expressed as 
\begin{align}
\int p(x, s=-1)\ell(-g(x))dx&=\int\frac{1-\sigma}{\sigma}p(x, s=+1)\ell(-g(x))dx \nonumber \\
&=\int\frac{1-\sigma}{\sigma}p(s=+1)p(x|s=+1)\ell(-g(x))dx \nonumber \\
&=(\pi + \rho)R^-_{s=+1}\left(\frac{1-\sigma}{\sigma}g\right). \nonumber 
\end{align}
\subsection*{Acknowledgment}
This work was supported by JSPS KAKENHI Grant Number 19K11858.


\end{document}